\documentclass[conference]{IEEEtran}
\usepackage{cite}
\ifCLASSINFOpdf
   \usepackage[pdftex]{graphicx}
\else
\fi
\usepackage[cmex10]{amsmath}
\usepackage{array}
\usepackage{url}
\usepackage{multirow}

\begin{document}
\title{Tunable and Growing Network Generation Model with Community Structures}

\author{\IEEEauthorblockN{Muhammad Qasim Pasta}
\IEEEauthorblockA{Karachi Institute of Economics and Technology\\
Karachi, Pakistan\\
Email: mqpasta@pafkiet.edu.pk}

\and
\IEEEauthorblockN{Zohaib Jan}
\IEEEauthorblockA{Shaheed Zulfikar Ali Bhutto Institute of  Science and \\ Technology\\
Karachi, Pakistan\\
Email:zohaib.jan@szabist.edu.pk }

\and
\IEEEauthorblockN{Arnaud Sallaberry}
\IEEEauthorblockA{LIRMM, Universit\'e Montpellier 3,\\
Montpellier, France\\
Email: arnaud.sallaberry@lirmm.fr}

\and
\IEEEauthorblockN{Faraz Zaidi}
\IEEEauthorblockA{Karachi Institute of Economics and Technology\\
Karachi, Pakistan\\
Email: faraz@pafkiet.edu.pk}
%\and
%\IEEEauthorblockN{James Kirk\\ and Montgomery Scott}
%\IEEEauthorblockA{Starfleet Academy\\
%San Francisco, California 96678-2391\\
%Fax: (888) 555--1212}
}


%Telephone: (800) 555--1212, Fax: (888) 555--1212}


\maketitle

\begin{abstract}

Recent years have seen a growing interest in the modeling and simulation of social networks to understand several social phenomena. Two important classes of networks, small world and scale free networks have gained a lot of research interest. Another important characteristic of social networks is the presence of community structures. Many social processes such as information diffusion and disease epidemics depend on the presence of community structures making it an important property for network generation models to be incorporated. 

In this paper, we present a tunable and growing network generation model with small world and scale free properties as well as the presence of community structures. The major contribution of this model is that the communities thus created satisfy three important structural properties: connectivity within each community follows power-law, communities have high clustering coefficient and hierarchical community structures are present in the networks generated using the proposed model. Furthermore, the model is highly robust and capable of producing networks with a number of different topological characteristics varying clustering coefficient and inter-cluster edges. Our simulation results show that the model produces small world and scale free networks along with the presence of communities depicting real world societies and social networks. 
%** chk if some other types of graph characteristics can be modified

\end{abstract}

%\keywords{Complex Networks \and Community Structures \and Network Generation Models \and  Scale Free Networks \and Small World Networks}

% \fi
\IEEEpeerreviewmaketitle

\section{Introduction}

Traditionally graph and network studies were made using regular and random graphs \cite{erdos59} until the late 1990's when two ground breaking discoveries were made about real world networks. The presence of low average path lengths and high clustering coefficients lead to the discovery of small world networks \cite{watts98} and the study of degree distribution following power-law lead to the discovery of scale free networks \cite{barabasi99}. These topological characteristics are commonly present in many real world networks such as social networks \cite{watts03}, biological networks \cite{dorogovtsev02} and information networks \cite{newman03}.

The discovery of small world and scale free properties have catalysed the research in the area of developing new graph and network generation models as networks with these properties appear readily across different and contrasting domains. This research area presents interesting challenges and new horizons for researchers to develop theories, models and algorithms based on the simulation and modeling of networks with domain dependent as well as domain independent characteristics. Furthermore, these models help us understand the underlying processes and structural changes taking place in many diverse real world networks.

An important characteristic of these networks is the presence of community structures. Networks in general and social networks in particular highly depend on a society based structure where groups of people are very well connected to each other and sparsely connected to people from other groups \cite{gilbert11}. This phenomena has been observed in networks from many different domains such as computer networks \cite{lim06}, biological networks \cite{wagner00} and maritime transportation networks \cite{ducruet12}. A number of network studies depend on the underlying community structure present in a network \cite{eriksen03}. For example, \cite{lambiotte09} studied communities of scientists and the role of information diffusion in the creation of knowledge. \cite{becker98} studies the effects of communities on the immunity coverage required to prevent disease epidemics in societies. Tunable methods that can generate networks with desired network characteristics and the presence of communities can be very useful in such studies as they can provide a benchmark for empirical evaluation. Researchers \cite{lancichinetti08,lancichinetti09} have proposed different models to generate networks with community structures in an attempt to generate networks which are topologically similar to real world networks.

An important feature often overlooked by different network generation models with community structures is the topological structure of a community itself. Usually models propose increased intra-community links and reduced inter-community links but the connectivity within a community is ill defined or follow only power-law degree distribution. The major contribution of the model is that we maintain three important topological characteristics within each community:
\begin{itemize}
\item The degree distribution of nodes follows a power-law.
\item The clustering coefficient is high.
\item Each community can be further divided into sub-communities, i.e.\ there are hierarchical communities in the network 
\end{itemize}

Each of the above characteristics has associated social semantics. Consider a co-authorship network where researchers collaborate to author manuscripts. Each research group represents a community as members regularly collaborate to increase intra-cluster edges which in turn results in high clustering coefficient. Subsequently, this research group also belongs to the community of researchers working in the same area across different research labs and different countries. These researchers collaborate less frequently but have still more edges when compared with research groups working in different domains. This creates a hierarchical community structure in the co-author network as argued by other researchers as well \cite{girvan02}. Each research group is usually headed by a senior professor with research publications, which means that a senior professor will have a high number of co-authors. Structurally this implies that the senior professor will have many edges connecting it to many authors. Usually each research group has a few senior researchers with high publication profile and the rest of the team comprises of researchers with low publications. These researchers are often associated with the senior professors while authoring an article justifying that every community demonstrates scale invariance power-law. Similarly the group dynamics and synergy is reflected by the people within a research group collaborating to author manuscripts. This results in a high number of triad formation which in turn results in high clustering coefficient for members of a research groups depicting a community. 

The contribution of this paper is that we propose a new network generation model. The proposed model is inspired by \cite{holme02} to generate networks with small world and scale free properties where we modify it to introduce community structures. The model caters the three described features present in community structures which is fundamental to many real world networks and specially in the case of social networks. Parameters to control inter-cluster connectivity and triad formation gives us more flexibility over the generation process and thus enables us to generate networks with desired properties. The networks produced using the proposed algorithm also exhibit small world and scale free properties. The model is tunable and robust as it can be used to generate a variety of networks by varying different parameters such as only scale free networks with community structures and networks with varying inter-cluster edges. 

The rest of the paper is structured as follows: In the next section, we review the literature related to network generation models. Section \ref{sec::proposed} describes the proposed model whereas section \ref{sec::discussion} analyses and explains the use of different parameters to generate networks with varying structural properties. Section \ref{sec::experimentation} presents the results of the evaluation of the networks generated using the proposed model satisfying the small world and scale free properties with clear community structures. Finally we conclude in section \ref{sec::conclusion} giving future research directions.

\section{Related Work}\label{sec::related}

We divide the literature review into two logical subsections.

\subsection{Models for Random, Small World and Scale Free Networks}

Earlier studies related to network models were focussed on generating random graph. Most notable work of all is the graph generation model by \cite{erdos59}. Molloy and Reed \cite{molloy95} proposed a model to generate graphs with desired degree distribution. Watts and Strogatz proposed the famous model to generate small world networks \cite{watts98} where the algorithm starts with a regular graph and random rewiring of edges based on some probability results in a small world graph with small average path length and high clustering coefficient. Albert and Barabasi introduced another important model \cite{barabasi99} based on preferential attachment to generate scale free networks.

Since the discovery of small world and scale free networks, a number of network models have been proposed to generate networks with these two properties. Most of these models are variants of the two basic models \cite{watts98,barabasi99} discussed above.
For example Holme and Kim \cite{holme02} introduce a triad formation step after the preferential attachment step in \cite{barabasi99} which creates triads in the network increasing the overall clustering coefficient. Other variants such as \cite{dorogovtsev02,guo05,fu06,klemm02,catanzaro04,wang08b,li12a} produce networks by introducing triads one way or the other and nodes connect using the preferential attachment rule to have a scale free degree distribution. 

Another approach for generating small world and scale free networks is the use of n-partite structure. Newman \textit{et al.}\cite{newman02a}  study a network generation model with arbitrary degree distribution. The goal is to generate affiliation networks similar (such as  co-authorship network\cite{newman03}) using random bipartite graphs. Guillaume and Latapy \cite{guillaume04} also used a similar idea as they identify bipartite graphs as an underlying structure for networks with small world and scale free properties. Bu \textit{et al.}\cite{bu07} used a n-partite structure, which is simply a generalization of the earlier proposed models. Good references on network generation models can be found in \cite{newman00,badham10,zaidi13a}.

\subsection{Models for Networks with Community Structures}

Li and Chen \cite{li06a} introduced a model for weighted evolving networks with community structures. The model incorporated three types of power-law distributions, first on the node degree, second on link weights and third on node strengths along with the presence of clear communities. The model does not produce networks with high clustering coefficient as nodes within a community do not follow triadic closure property.

Xie \textit{et al.} \cite{xie07} proposed a community-based evolving network model where they focus on the cumulative distribution of community sizes which also follows power-law in real world networks. As a result, when new connections between communities are added, or a new node to an existing community is added, communities with larger sizes are selected preferentially.

Zhou \textit{et al.} \cite{zhou08} identify two important topological characteristics, first, intra-cluster connections are very dense as compared to inter-cluster connections and second, size of communities often follows a power-law just as \cite{xie07} proposed. Based on these characteristics, they propose a weighted growing model with power-law distributions of
community sizes, node strengths, and link weights. % can't find a difference except for weight

Kumpula \textit{et al.} \cite{kumpula09} utilize the concepts of cyclic closure and focal closure from sociology to propose a model to generate a weighted network with communities. New links are created preferably through strong ties which make these links more stronger. The model also allows the removal of nodes to mimic real world scenarios where nodes may leave a network.

Xu \textit{et al.} \cite{xu09} introduce a model with communities that gives a realistic description of local events using three processes, adding new intra-community nodes, new intra-community links or new inter-community links. The model uses preferential attachment mechanism resulting in power law degree distribution but since the intra-community links only connect on the basis of node degree, the network lacks triads, producing networks with low clustering coefficients.

Lancichinetti and Fortunato \cite{lancichinetti09} propose an algorithm to generate benchmarks to test clustering algorithms for directed/undirected and weighted/unweighted graphs with optional overlapping communities. This algoritm produces networks following power-law distribution for node degree as well as community sizes. They do not address the internal structure of each community as we do in this paper.

Badham and Stocker \cite{badham10} propose a spatially constructed algorithm to generate networks with tunable degree distribution, clustering coefficient and assortativity with the objective that such models should be flexible to generate networks with varying values of these properties giving more control over the generation process. They do not explicitly include the generation of community structures in their model.

Ren \textit{et al.} \cite{ren12} study the connecting patterns among existing papers in co-authorship networks and highlight that existing models cannot correctly model high clustering in such networks. Their proposed model can generate networks with power-law degree distribution, high clustering coefficient and the size distribution of co-citation clusters as observed in co-authorship networks.

Moriano and Finke \cite{moriano13} also propose a model with small world and scale free properties along with groups of nodes densely connected to each other and sparsely connected with other nodes. The model helps to explain networks with extended power law degree distributions and clustering coefficient that does not diminish as the size of the network grows very large. The connectivity of new nodes probabilistically chooses nodes of same type or different type to form community structures.

Zaidi \cite{zaidi13a} proposed a model to generate clustered small world networks. The author first demonstrates that small world networks can be produced from completely random graphs by introducing a little order in them, which is a contrasting approach to the famous model of \cite{watts98}. The further extends this model to generate clustered networks with small world properties where communities connect randomly to other communities. The model does not generate scale free networks.

Zaidi \textit{et al.} \cite{zaidi13b} also proposed a static network generation model with community structures i.e\ nodes added at the start remain the same throughout the algorithm and only edges are rewired to create communities. The model is probabilistic and increases the edge connectivity among nodes closer to each other and reduces edges among nodes far apart in the network. But the process is not a growing one, and it is not parametrized to generate desirable clustered network as compared to the model proposed in this study. Furthermore this model does not ensure the presence of the three structural properties for a community discussed in the introductory section of this paper.

All the different models for generating networks with community structures discussed above do not focus on the internal structure of communities just as we do in this paper. We focus on three structural properties present in our society and propose a model to simulate these properties. 

\begin{figure}[t]
\begin{center}
\includegraphics[width=0.2\textwidth]{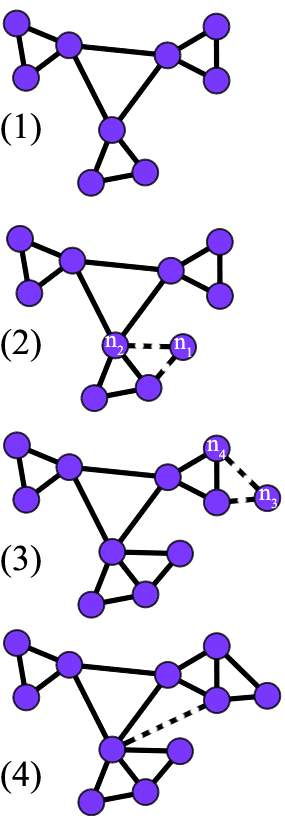}
\end{center}
\caption{Steps of the proposed model. (1) Step \ref{ctriads} with \textit{c}=3 triads where every triad is connected to every other triad. (2) Step \ref{newnode1}, A new node $n_1$ is added and forms a triad with probability $P_t$ with a neighbor of $n_2$. (3) Step \ref{newnode2}, A new node $n_3$ is added and forms a triad with probability $P_t$ with a neighbor of $n_4$. (4) Step \ref{pc}, The communities of newly added nodes $n_1$ and $n_3$ which are selected on the basis of preferential attachment forms an inter-cluster edge with probability $P_c$.}
\label{fig::modelsteps}
\end{figure}

\section{Proposed Model}\label{sec::proposed}

The model takes as input the desired number of nodes in the network $n$, the number of edges for each newly added node $m$, the minimum number of communities $c$, the probability of triad formation $P_t$ and probability $P_c$ of having inter-cluster edges. The model starts with an empty network. Rest of the steps are explained below:
\begin{enumerate}
%\item \label{start} Start with an empty network without any nodes and edges.
\item \label{ctriads} Add $c$ triads representing $c$ communities in the network. Each node in a triad belongs to the community of the triad. Every community thus created is then connected to every other community. An edge is created between randomly selected nodes from different communities. This step generates a graph as shown in figure \ref{fig::modelsteps}(1).
\item \label{newnode1} Add a new node $n_1$ and connect it to an existing node $n_2$ which is selected on the basis of preferential attachment. Now $n_1$ belongs to the community of $n_2$ as shown in figure \ref{fig::modelsteps}(2).
\item \label{pt1} With probability $P_t$, either $n_1$ connects to $m-1$ preferentially selected neighbors of $n_2$ belonging to the same community as $n_2$ forming a triad or $n_1$ connects to other nodes preferentially selected from the community of $n_2$ (which might not necessarily be a neighbor of $n_2$). Figure \ref{fig::modelsteps}(2) shows that the newly added node connected to one of the neighbors of $n_2$.
\item \label{newnode2} Add a new node $n_3$ and connect it to an existing node $n_4$ which is again selected on the basis of preferential attachment. Now $n_3$ belongs to the community of $n_4$. We make sure that $n_3$ does not belong to the community of the node added in the previous node-addition step to the network. Figure \ref{fig::modelsteps}(3) shows this step.
\item \label{pt2} With probability $P_t$, either $n_3$ connects to $m-1$ preferentially selected neighbors of $n_4$ belonging to the same community as $n_4$ forming a triad or $n_3$ connects to other nodes preferentially selected from the community of $n_4$ ( which might not necessarily be a neighbor of $n_4$), as shown in Figure \ref{fig::modelsteps}(3).
\item \label{pc} With probability $P_c$, add an edge between two preferentially selected nodes belonging to the two communities to which nodes were added in the previous steps, which is the communities of $n_2$ and $n_4$ as shown in figure \ref{fig::modelsteps}(4).
\item \label{repeat} Repeat from step \ref{newnode2} until number of nodes in the network becomes $n$.

\end{enumerate}

\begin{table}
\begin{tabular}{|l|c|c|c|c|}
\hline 
S.No & Number     & Initial Number  & Probability of    & Probability of  \\
Key  & of Nodes   & of Triads       & Triad Formation   & Inter-Cluster Edges\\
\hline \hline
1 & 1000 & 10 & 0.5 & 0.01 \\
\hline
2 & 1000 & 10 & 0.5 & 0.10 \\
\hline
3 & 1000 & 10 & 0.5 & 0.5 \\
\hline
4 & 1000 & 10 & 0.5 & 1.0 \\
\hline
5 & 1000 & 10 & 1.0 & 0.01 \\
\hline
6 & 1000 & 10 & 1.0 & 0.10 \\
\hline
7 & 1000 & 10 & 1.0 & 0.5 \\
\hline
8 & 1000 & 10 & 1.0 & 1.0 \\
\hline
9 & 1000 & 20 & 0.5 & 0.01 \\
\hline
10 & 1000 & 20 & 0.5 & 0.10 \\
\hline
11 & 1000 & 20 & 0.5 & 0.5 \\
\hline
12 & 1000 & 20 & 0.5 & 1.0 \\
\hline
13 & 1000 & 20 & 1.0 & 0.01 \\
\hline
14 & 1000 & 20 & 1.0 & 0.10 \\
\hline
15 & 1000 & 20 & 1.0 & 0.5 \\
\hline
16 & 1000 & 20 & 1.0 & 1.0 \\
\hline
17 & 10000 & 10 & 0.5 & 0.01 \\
\hline
18 & 10000 & 10 & 0.5 & 0.10 \\
\hline
19 & 10000 & 10 & 0.5 & 0.5 \\
\hline
20 & 10000 & 10 & 0.5 & 1.0 \\
\hline
21 & 10000 & 10 & 1.0 & 0.01 \\
\hline
22 & 10000 & 10 & 1.0 & 0.10 \\
\hline
23 & 10000 & 10 & 1.0 & 0.5 \\
\hline
24 & 10000 & 10 & 1.0 & 1.0 \\
\hline
25 & 10000 & 20 & 0.5 & 0.01 \\
\hline
26 & 10000 & 20 & 0.5 & 0.10 \\
\hline
27 & 10000 & 20 & 0.5 & 0.5 \\
\hline
28 & 10000 & 20 & 0.5 & 1.0 \\
\hline
29 & 10000 & 20 & 1.0 & 0.01 \\
\hline
30 & 10000 & 20 & 1.0 & 0.10 \\
\hline
31 & 10000 & 20 & 1.0 & 0.5 \\
\hline
32 & 10000 & 20 & 1.0 & 1.0 \\
\hline
\end{tabular}
\caption{The table shows the 24 different combinations possible for 4 parameters and different possible values. The serial number will be used as a key to identify networks and their parameters used.} \label{tbl::parameters}
\end{table}

\section{Discussion}\label{sec::discussion}

As described above, the model uses five parameters, $n,m,c,P_t,P_c$. The parameter $n$ defines the number of nodes desired in the final network and $m$ defines the number of connections each newly added node will have in the network (except for the nodes added in the initial triads in step \ref{ctriads}). 

%\textbf{Number of edges $m$ for each node:} This parameter can be used to control the node-edge density as with the increase of this parameter, the average degree of each node increases. The number of edges for the entire network can be calculated using this parameter with the following equation:

%$$ e=\{((n-(c*3))*m)+1\} + \{(c*3)+c\}  $$

%The first term in the above equation represents the number of nodes added to the network after step \ref{ctriads} times the number of edges for each node, and the second term represents the number of edges for the initially added triads in the network.

\textbf{Preferential Attachment for individual nodes:} The probability of a new node preferentially selecting an existing node $n_i$ from the set of current vertices $V$ is a function of the degree of node $n_i$ which can be calculated using:

$$ P(n_k)= \frac{degree(n_k)}{\sum{degree(n_j)}} ,\forall j \in V $$

This ensures that the nodes are selected based on preferential attachment. 

\textbf{Preferential Attachment for each community:} The probability that a new node selects a community $c_k$ to attach can be estimated as:

$$ P(c_k)=\frac{\sum degree(n_i)}{\sum{degree(n_j)}} ,\forall i \in c_k,  \forall j \in V  $$

As the network grows, the community sizes vary as a function of the high degree nodes present in that community. This ensures that communities of different sizes evolve in the network where the degree distribution of community sizes follow scale free behavior. This is because new nodes select a node to attach based on its degree, which in turn implies that a community is selected based on preferential attachment of nodes present in a community as shown in the above equation.

\textbf{Minimum Number of Communities $c$:} This parameter controls the minimum number of communities we want to generate in the network. Further Communities and sub-communities form as order emerges from the connectivity of new nodes entering the network probabilistically. As the number of nodes increases, sub-communities increase depicting the natural evolution process of communities in real world networks. A simple variation for this parameter would be to use the value 1 signifying only 1 community, along with the triad formation step using $P_t$, the behavoir of the network would be the same as the model proposed by \cite{holme02}. Small sub-communities will still form in this network but they will not be clearly separable. Another important variation would be if we use $c=1, m=1$ and $P_t=0$ i.e.\ eliminating the triad formation step, the model generates random scale free networks similar to \cite{barabasi99}.

\textbf{Probability of Triad Formation $P_t$:} This parameter controls the presence of triads in the network. The triad formation step is performed with a probability $P_t$, or a preferential attachment step is performed with probability $1-P_t$ instead of triad formation step. Both the triad formation step or preferential attachment step, the new node is only connected to nodes from the same community. A value of $P_t=0$ means that in steps \ref{pt1} and \ref{pt2}, triads are not formed, as a result of which the overall clustering coefficient remains quite low. The network thus generated is a random scale free network with communities. A value of $P_t=1$ would mean that every node added to this network with $m$ edges, forms triads for every edge making the overall clustering coefficient quite high as is the case for small world networks.

\textbf{Probability of Inter-cluster Edges $P_c$:} The inter-cluster density is controlled through this probability. A value of $P_c=0$ means that no further intra-cluster edges would be added to the network as described in step \ref{pc}. This results in well separated communities with exactly two intra-cluster edges for each community which were added in step \ref{ctriads}. These edges are added so that the final network obtained, remains a connected network. A value of $P_c=1$ results in high inter-cluster edges making it difficult to distinguish communities structurally.

We demonstrate the effects of varying these parameters empirically in the next section as we generate numerous networks using the proposed model.

%\section{Analysis}\label{sec::analysis}
%use zhou08 for mathematical analysis 

\begin{figure}
\begin{center}
\includegraphics[width=0.4\textwidth]{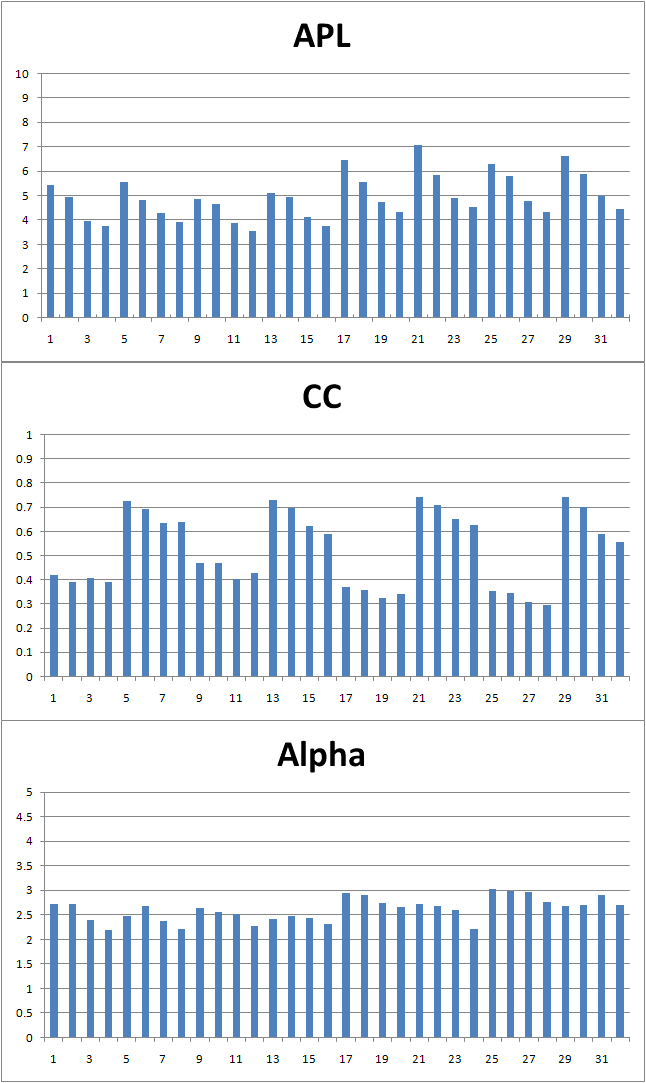}
\end{center}
\caption{Different metrics calculated for the 32 generated graphs showing that the graphs are indeed small world and scale free. (a)APL=Average Path Length (b) CC=Clustering Coefficient (c)) Alpha=Power-law coefficient.}
\label{fig::metrics}
\end{figure}

\begin{figure}
\begin{center}
\includegraphics[width=0.4\textwidth]{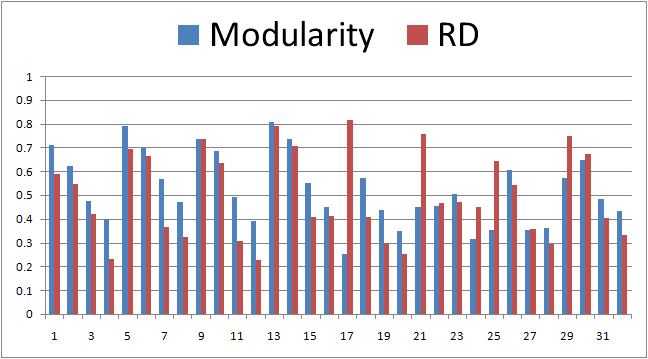}
\end{center}
\caption{Values of Modularity (Q)-Blue Bar and Relative Density (RD)-Maroon Bar, obtained after running a clustering algorithm on the generated networks. High values suggest the presence of community structures in the generated network.}
\label{fig::qrd}
\end{figure}

%** replace this with running times
\begin{figure}
\begin{center}
\includegraphics[width=0.4\textwidth]{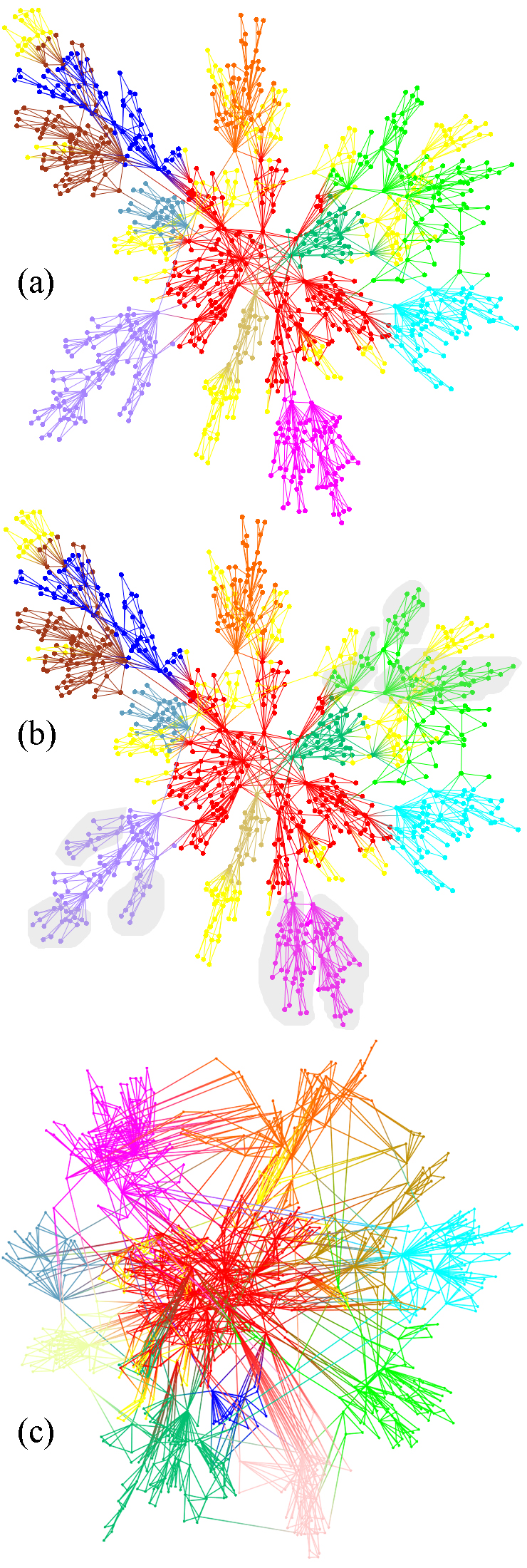}
\end{center}
\caption{Figure shows the graphical representation of the networks generated using the proposed model. (a) Network 5 with clearly seperated communities. The top 10 communities in terms of size are uniquely colored and smaller communities are colored in yellow. (b)Network 5 with smaller communities outlined within larger communities. (c) Network 6 with more inter-cluster edges and less seperation between communities.}
\label{fig::samples}
\end{figure}

\section{Experimentation and Results}\label{sec::experimentation}

We tested the proposed model with the parameter values of $n=\{1000,10000\}, c=\{10,20\}, P_t=\{0.5,1.0\}$ and $P_c=\{0.01,0.1,0.5,1.0\}$ which makes 32 different combinations. We used $m=\{2\}$ for all performed experiments which means that every new node entering the network has degree 2 to start with, which increases probablistically as other new nodes select previously added nodes as their connections. Table \ref{tbl::parameters} shows all these combinations and a key is assigned to each combination to uniquely identify a network and the parameters used to generate it. For each of these parameter values, we generated 5 networks each, and used the average of the metrics obtained as a result, eliminating potential outliers and exceptions that might bias the ultimate results as the algorithm is probabilistic in nature. 

Figure \ref{fig::metrics} show the values of average path length, average clustering coefficient and power-law coefficient(alpha) for all the generated graphs. This is to demonstrate that our graphs are indeed small world and scale free networks. All the graphs have an average path length between 3 and 7. Path lengths of around 7 are observed in networks where we introduce inter-cluster edges with a probability of  $0.01$. This results in clearly separated communities with very little inter-cluster edges, which in turn results in increased distances among nodes from different clusters. For the clustering coefficient values, we used two parameter values, $0.5$ where triads are formed only for $50\%$ newly added nodes with $m$ edges giving CC values in the range of 0.3 and 0.5, and the parameter value $1.0$ where all nodes and edges added to the network belong to at least one triad which raises the CC around $0.65$. Finally for the alpha value, since all our connectivity is based on preferential attachment, all the 32 generated networks have values between 2 and 3.

In order to show the presence of community structures in the generated networks, we used two well known metrics, Modularity (Q)\cite{newman04} and Relative Density \cite{schaeffer07}. We clustered the generated graphs using the method proposed by Newman \cite{newman06} which generates flat clusters. We calculated the Q and RD values which are shown in figure \ref{fig::qrd}. Consistently high values clearly demonstrate the presence of community structures present in the networks. As the inter-cluster edges are increased, both Q and RD values decrease implying that the control parameter $P_c$ can be used to generate communities with low or high inter-cluster edges which subsequently affects the Q and RD values.

Figure \ref{fig::samples}(a,c) are two sample networks (Network 5 and 6 in Table \ref{tbl::parameters}) generated from the proposed model. Both these networks show community structures with different color encodings for the top 10 communities in terms of size. More communities with smaller sizes are encoded with yellow color.

Figure \ref{fig::samples}(b) shows Network 5 where larger clusters contain clear separation and can be re-clustered to form hierarchical community structures. Some of these clusters are highlighted and can be visually compared with Figure \ref{fig::samples}(a).

\begin{figure}
\begin{center}
\includegraphics[width=0.4\textwidth]{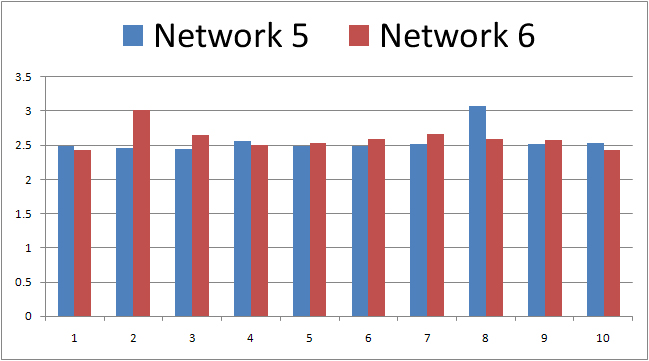}
\end{center}
\caption{The graph shows the values of power-law coefficient for the top 10 communities in terms of size for the two networks shown in \ref{fig::samples}. All the values are consistently between 2 and 3 aprrox. Network 5 has parameters: $n=1000$,$c=10$,$P_t=1.0$,$P_c=0.01$ and Network 6 with parameters:$n=1000$,$c=10$,$P_t=1.0$,$P_c=0.1$.}
\label{fig::alpha}
\end{figure}

\begin{figure}
\begin{center}
\includegraphics[width=0.4\textwidth]{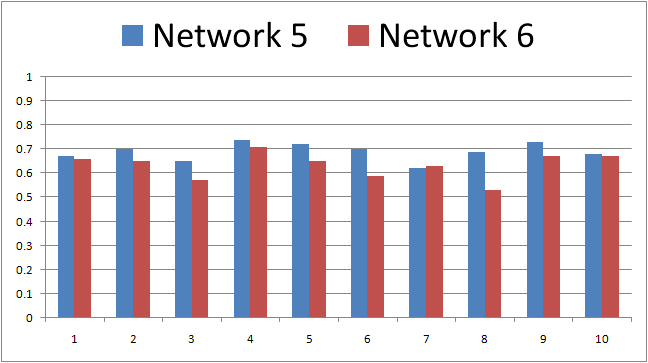}
\end{center}
\caption{The graph shows the values of clustering coefficient for the top 10 communities in terms of size for the two networks shown in \ref{fig::samples}. All the values are consistently above 5. Network 5 has parameters: $n=1000$,$c=10$,$P_t=1.0$,$P_c=0.01$ and Network 6 with parameters:$n=1000$,$c=10$,$P_t=1.0$,$P_c=0.1$.}
\label{fig::cc}
\end{figure}

To prove our claim that the communities thus produced follow power-law degree distribution with high clustering coefficient, we plotted the power-law coefficient and clustering coefficient of the top ten communities in terms of node-size for the two graphs presented in figure \ref{fig::samples}. Figure \ref{fig::alpha} shows the power-law coefficient which lies in the range [2,3] clearly showing that the communities thus produced follow scale free behavior. Similarly figure \ref{fig::cc} shows high clustering coefficient values for the biggest communities in networks 5 and 6 re-affirming that formation of triads ensures that communities have a large number of triads.

Finally to show the scalability of the proposed model, we plot the execution times for the generation of different size networks in Table \ref{tbl::time}. The proposed model has been implemented using Tulip graph library\footnote{The source code of the model can be requested from the first author through email.}\cite{auber03a}. The running times given in the table are on a standard intel core i5 machine with 4Gb Ram.

\begin{table}
\begin{center}
\begin{tabular}{|l|c|c|}
\hline 
S.No & Nodes     & Time in Seconds   \\
\hline \hline
1 & 1000   & below 1  \\
\hline
2 & 10000   & 2     \\
\hline
3 & 100000  & 224   \\
\hline
4 & 1000000 & 21966 \\
\hline
\end{tabular}
\end{center}
\caption{The table shows the running times in seconds for generating different size networks using the proposed model.} \label{tbl::time}
\end{table}

In terms of complexity of the proposed algorithm, the most complex task is the calculation of probability for a new node entering the network based on preferential attachment. This task, in the worst case requires $n*nc_{large}$ steps where $n$ is the desired number of nodes in the network and $nc_{large}$ is the number of nodes in the largest community. The preferential attachment probability for every node added to the network is calculated in the worst case, with every other node in the largest community. If a network is generated with only a single community, $nc_{large}$ becomes $n$ and the complexity of the entire algorithm would then become $O(n^2)$. For a network where the community sizes are sparsely distributed, as is the case with many real world networks, this complexity becomes $O(n*nc_{large})$ with $nc_{large} << n$.

%\begin{figure}[h]
%\begin{center}
%\includegraphics[width=0.45\textwidth]{images/time.jpg}
%\end{center}
%\caption{The graph shows the values of power-law coefficient for the top 10 communities in terms of size for the two networks shown in \ref{fig::samples}. All the values are consistently between 2 and 3 aprrox. Network 5 has parameters: $n=1000$,$c=10$,$P_t=1.0$,$P_c=0.01$ and Network 6 with parameters:$n=1000$,$c=10$,$P_t=1.0$,$P_c=0.1$.}
%\label{fig::time}
%\end{figure}

\section{Conclusion}\label{sec::conclusion}

In this paper, we have introduced a new tunable and growing network generation model which incorporates the well known small world and scale free properties as well as the presence of community structures. The model incorporates three important features of community structures from our society. First, the node degree distribution within a community follows power-law behavior, second, the clustering coefficient within communities is high and finally there is a hierarchical community structure within communities. The model is very flexible and robust and can be used to generate a variety of networks as per requirements. These characteristics can be very useful for generating benchmark and test datasets for empirical studies.
Although the model is flexible, but it does not include domain dependent knowledge and cannot be used to generated networks with  structural properties other than community structures, small world, scale free and random. This work can clearly be extended to generate networks for particular domains such as biological networks, computer networks. Further more, the current model does not simulate the dynamic changes like removal of previously added nodes or edges, changing previously added edges which are important characteristics for the recently studied social networks. We intend to extend this study towards this direction as well.


\bibliographystyle{IEEEtran}
\bibliography{visu}
%\begin{thebibliography}{1}
%\end{thebibliography}

\end{document}